\documentclass{article}
 \title{A Dynamic I/O Model for TRACON Traffic Management}

\author{
  Maxime Gariel\thanks{Graduate Research Assistant, School of Aerospace Engineering,
    maxime.gariel@gatech.edu, corresponding author},
  \  John-Paul Clarke\thanks{Professor, School of Aerospace
  Engineering, john-paul.clarke@ae.gatech.edu.}
  \ and Eric Feron\thanks{Professor, School of Aerospace
  Engineering, feron@gatech.edu.}\\
  {\normalsize\itshape
   Georgia Institute of Technology Atlanta, GA, 30332-0150, USA}\\
 }

\usepackage{subfigure}
\usepackage{graphicx}

\begin{document}

\maketitle

\begin{abstract} This work investigates the TRACON flow management around a major
airport. Aircraft flows are analyzed through a study of TRACON
trajectories records. Rerouting and queuing processes are
highlighted and airport characteristics are shown as function of the
number of planes in the TRACON. Then, a simple input-output TRACON
queuing and landing model is proposed. This model is calibrated and
validated using available TRACON data. It reproduces the same
phenomenon as the real system. This model is used to show the impact
of limiting the number of aircrafts in the TRACON. A limited number
of aircraft does not increase delays but reduces the controller's
workload and increases safety.
\end{abstract}

\section{Introduction}

Airspace, especially Terminal Radar Approach Control (TRACON) area
around major airports becomes more and more congested . As the
concentration of aircrafts is the highest, TRACON areas are very
critical areas for controllers. The objective of this paper is to
present a study about the dynamics of the traffic congestion in the
arrival process at a major airport. A motivation is to regulate the
flows of entering aircraft in the TRACON area in order to reduce air
traffic controller's workload and to avoid control tactical
difficulties. This is already the aim of the Center TRACON
Automation System (CTAS)\cite{CTAS} that provides automation tools
for planning and controlling arrival air traffic. Many traffic
control techniques and tools are applied to reduce traffic
congestion. Ground Delay Program (GDP) \cite{GDP} is used to
decrease the rate of incoming flights into an airport by delaying
takeoffs, when it is projected that arrival demand will exceed
capacity. For en-route control, Traffic Management Advisor (TMA)
\cite{TMA} and Multi-Center Traffic Management Advisor (McTMA)
\cite{McTMA} are used to control arriving aircraft that enter the
Center from an adjacent Center or depart from feeder airports within
the Center. Those systems have already proven their capabilities
\cite{McTMAresults}. TRACON area control can use programs such as
Descent Advisers (DA) or Final Approach Spacing Tool (FAST)
\cite{FAST}. Finally, ground operations and efficient runway
operation planning \cite{Ioannis} can also improve landing
capacities. The aim of those different layers of traffic control is
to ensure an arrival flow as smooth as possible. We are interested
in adjusting the density in the TRACON area and optimizing airports
landing capacities by regulating the incoming traffic flow.
Theoretically, runway capacities can be calculated using analytical
formulas and nominal separation standards. In this paper, those
capacities are determined and analyzed through available TRACON
data. Then, we propose an experimental queuing model to highlight
airport performance function of the demand. The airport performance
is the effective landing capacity, which depends on the number of
runway in use. This arrival model is based on a similarly model
developed by Nicolas Puget in \cite{pujet}. The model proposed is
validated with San Francisco TRACON records. Finally, the model is
used to show that limiting the number of aircrafts in the TRACON
does not increase delays but reduces the controller's workload and
increases safety.

\section{Data Presentation and Analysis}\label{section:dataPresentation}
\subsection{Data Source}
The data used in this article were provided by San Francisco
International Airport Noise Abatement Office. The data are the
records of all the planes in the San Francisco Bay TRACON for the
first 3 semesters of 2006.

\subsection{Definitions}
\begin{figure}[htbp]
\begin{center}
\includegraphics[width = 0.5\textwidth]{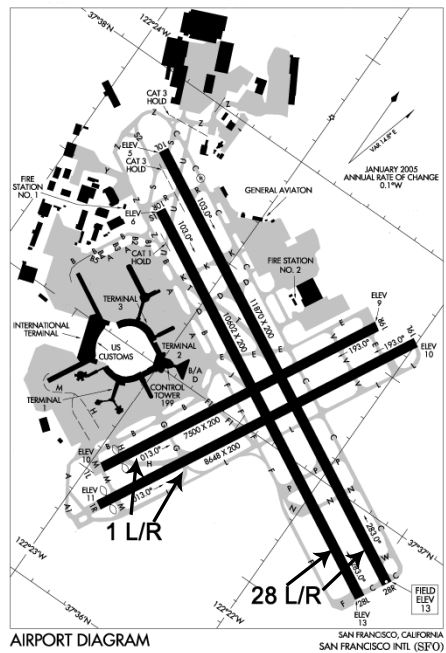}
\caption{San Francisco International Airport
diagram}\label{fig:sfoconfig}
\end{center}
\end{figure}

The focus is made on the arrival process. Hence, in this article,
the term \emph{system} will refer to the TRACON area which is a 50
Nm radius circle centered at Oakland Airport. The \emph{inputs} to
the system are the planes entering the TRACON, intending to land at
SFO airport on 28L/R runways (figure \ref{fig:sfoconfig}), and the
\emph{outputs} are the planes actually land ing on 28L/R runways at
SFO airport. 28R and 28L runway are only 750ft apart which does not
allow simultaneous parallel automatic landings. Hence, if the
weather is not good enough for sight landing, only one runway can be
used.

\begin{figure}[htbp]
\begin{center}
\includegraphics[width = 0.5\textwidth]{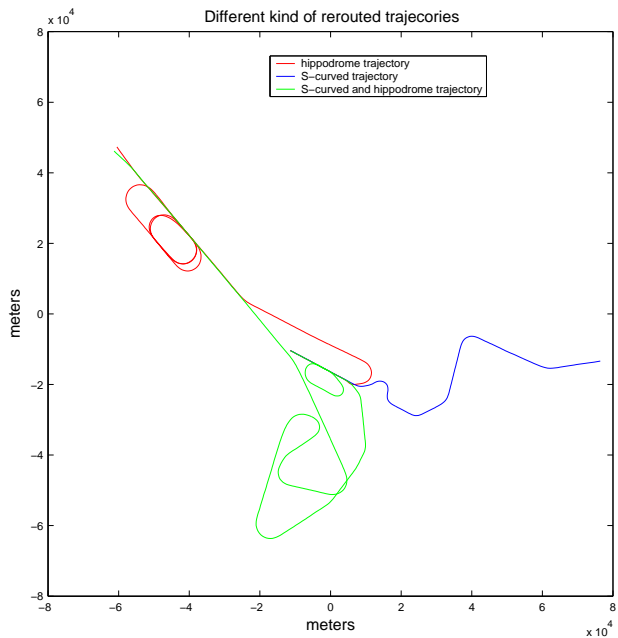}
\caption{Different kind of rerouted trajectories}
\label{fig:DifferentKindOfRerouting}
\end{center}
\end{figure}

Different parameters can be used to define the behavior of the
system. The evolution of the outputs (planes landing) can be
analyzed as a function of the inputs (planes entering the TRACON)
and a function of parameters such as the number of runways in use.
But we can also characterize it by the shape of planes trajectories
with in the same parameters. Each plane has a different track
leading to the runway. The analysis is made on 2D trajectories. For
the analysis purpose, tracks can be defined as \emph{straight} or
\emph{rerouted}. A \emph{straight} track corresponds to the
trajectory of a plane that could go directly from its entering point
in the TRACON to the  runway threshold just following the way
points. A \emph{rerouted} track corresponds to the trajectory
 of a plane that can not land directly and has to  change its
 trajectory. It does not follow the nominal track and has to make
 a curved trajectory or even hippodromes (figure \ref{fig:DifferentKindOfRerouting}).

\subsection{Data selection}
\begin{figure}[htbp]
\begin{center}
\includegraphics[width = 0.5\textwidth]{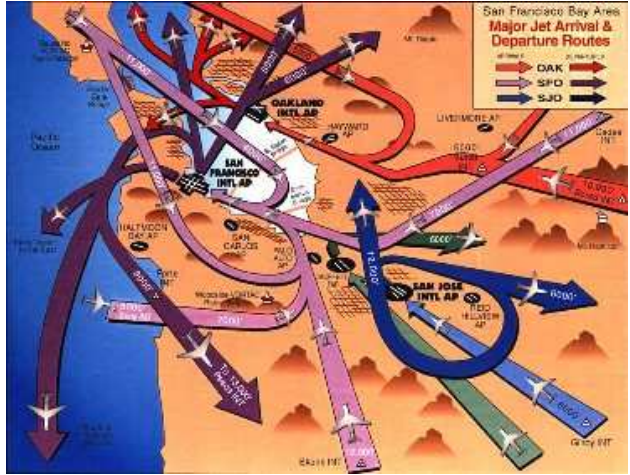}
\caption{West Configuration in San Francisco Bay}
\label{fig:westConfiguration}
\end{center}
\end{figure}

The analysis presented in this paper
focuses on the arrivals at San Francisco Airport. The presence of
several airports (Oakland Intl Ap, San Francisco Intl At, San Jose
Intl AP and many small airports) in a small perimeter such as San
Francisco Bay implies dedicated routes for landing and taking off
for each airport in the TRACON (figure \ref{fig:westConfiguration}).
Those routes decrease the degrees of freedom to reroute planes.
Moreover, for practical reasons such as fuel capacity and for
financial reasons, it is required to minimize the number of waiting
aircrafts and their waiting time. San Francisco International
Airport (figure \ref{fig:sfoconfig}) has four runways arranged in
two sets of parallel runways. The terrain
 configuration (presence of residential areas at west, north and south
 of the airport) is such that the ``West'' configuration is the
most used in order to reduce the noise over residential areas.
Actually, planes take off on 01L/R runways, departing over the bay
and land on 28L/R runways, maximizing the distance over the water in
order to reduce noise. We will use the planes landing in this
configuration (i.e on 28L and 28R runways).

The routes for landings and takeoffs are well defined, hence we can
assume there is no correlation between them: the planes flying to
SFO do not interact with planes flying from or to another airport or
just transiting.

The configuration of SFO airport is such that departures and
landings are on separate runways. The runways are crossing but we
will assume that taking off planes do not have influence on landing
planes.

\subsection{Radar track analysis}
\begin{figure}[ht]
\centering \subfigure[\emph{Direct} tracks]{\label{fig:tracks-a}
    \includegraphics[width = 0.4\textwidth]{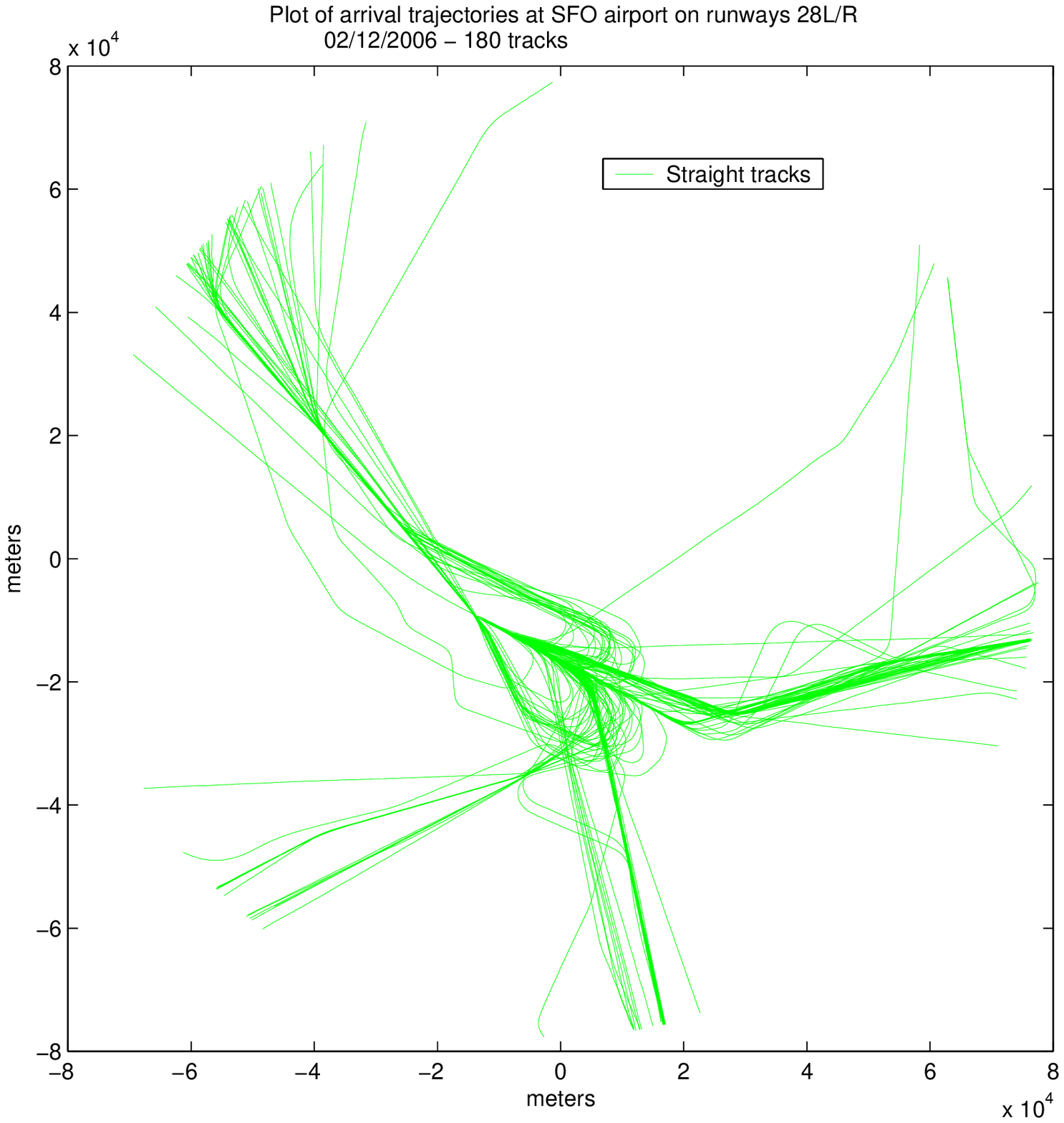}}
\subfigure[\emph{Rerouted} tracks]{\label{fig:tracks-b}
    \includegraphics[width = 0.4\textwidth]{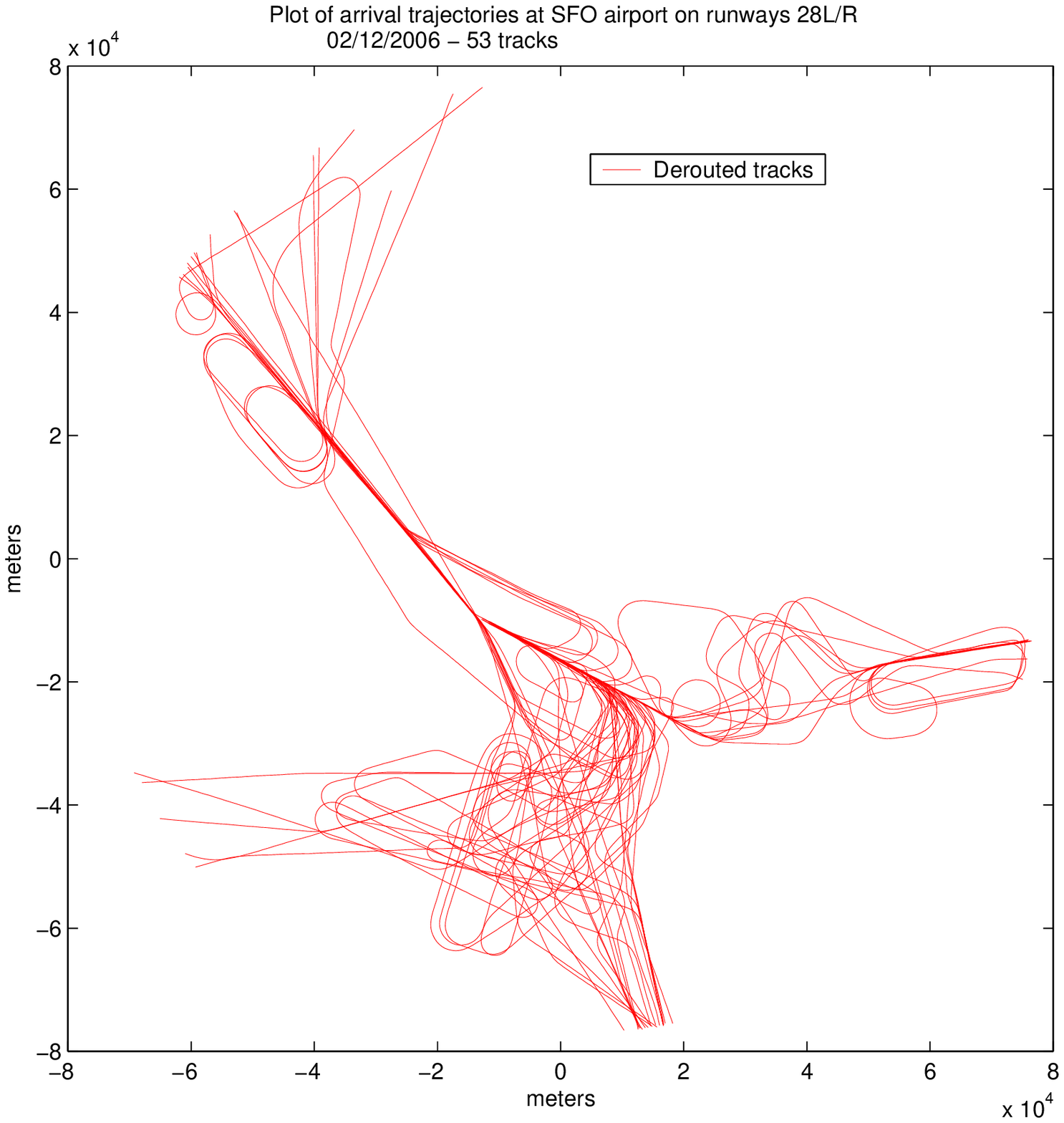}}
\caption{SFO tracks analysis} \label{fig:tracks}
\end{figure}

Each track is analyzed in order to determine whether it is a direct
track or a rerouted one. An algorithm has been developed to identify
rerouted tracks from direct tracks. The figure \ref{fig:tracks}
presents the results of this algorithm on a given day of arrivals at
SFO.\newline

Figure \ref{fig:tracks-a} presents the tracks identified as direct.
Almost all the plotted tracks are pretty direct to the runway,
considering the routes and way points that the airplanes must
follow. Figure \ref{fig:tracks-b} presents the tracks identified as
rerouted. All those tracks contain a change of direction not
corresponding to the route the airplane should follow for a nominal
approach. The rerouting is done either by \emph{smooth} changes of
trajectory, e.g \emph{S-turns}, or either by important changes of
trajectory, such as hippodromes (Figure
\ref{fig:DifferentKindOfRerouting}).

\subsection{Data Analysis}\label{subsection:dataanalysis} In this section, we investigate the
evolution of the system in two different ways. One is time-based
analysis and the other is aircraft-based.

\subsubsection{Time based methodology} In this
methodology, we divide the time in periods of fixed length during
which we will count:
\begin{itemize}
\item The number of planes present in the system which corresponds to
the number of planes that are, at some point of this period, in the
TRACON area (figure \ref{fig:timebasedanalysis}).
\item The number of planes entering the system during this period.
\item The number of planes landing during this period.
\item The number of planes rerouted during that period.
\end{itemize}
It is to note that a same plane can be counted as present and
rerouted on several time periods but enters or lands only once.

\subsubsection{Time based evolution of the system} 
\begin{figure}[ht]
\begin{center}
\includegraphics[width = 0.6\textwidth, angle = -90]{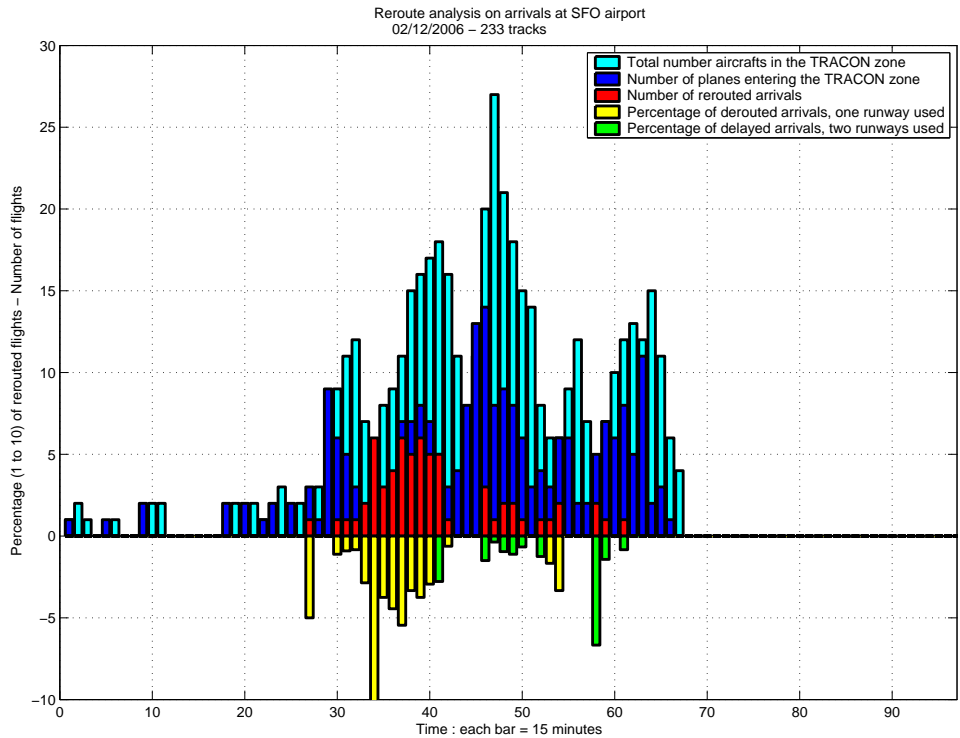}
\caption{Arrivals analysis February the 12th
2006}\label{fig:dailyAnalysis}
\end{center}
\end{figure}

Figure
(\ref{fig:dailyAnalysis}) presents the results of the time based
analysis. The time is on the abscises axis. The positive ordinates
axis is a number of flight while the negative ordinates axis
represents the proportion of the total number of planes in the
system that is rerouted, scaled from 1 to 10. Each bar stands for a
15 minutes time period. The light blue bars represent the total
number of aircrafts, the dark blue bars represent the number of
planes entering the system, the red bars represent the number of
rerouted planes. Yellow indicates that only one runway is in use
while green stands for 2 runways.

As shown in figure \ref{fig:dailyAnalysis}, in case of one runway in
use, the number of rerouted aircrafts increases with the number of
planes in the system. As soon as the second runway opens, the number
of rerouted aircrafts drops. In the two runways configuration, even
a pick of incoming aircrafts or a large number of aircrafts present
in the TRACON does not increase significantly the number of rerouted
planes. This methodology is more developed and used in
\cite{McTDMABenefits}.

\subsubsection{Aircraft based methodology}\label{sec:aircraft
methodology def} The evolution presented in this section is based on
a ``plane'' point of view. We look at the flow of entering
aircrafts, the flow of landing aircrafts, the number of planes
present in the system, the number of planes rerouted. Those terms
need to be well defined. To calculate flows, a time period $T$ is
required.  The term ``for each aircraft'' means that the count has
been done for all the aircrafts contained in the data used. The
total amount of selected aircraft is around 95,000.

\begin{figure}[ht]
\centering \subfigure[Time-based methodology to count the number of
plane present in the TRACON]{\label{fig:timebasedanalysis}
    \includegraphics[width = 0.7\textwidth]{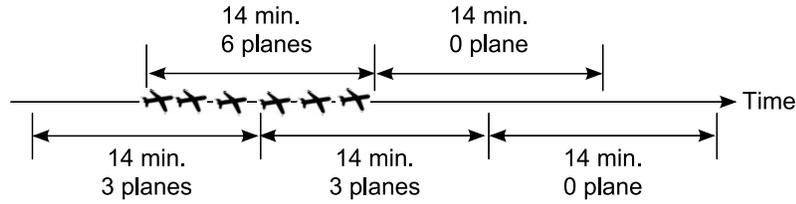}}
\subfigure[Aicraft-based methodology to count the number of plane
present in the TRACON]{\label{fig:aircraftbasedanalysis}
    \includegraphics[width = 0.7\textwidth]{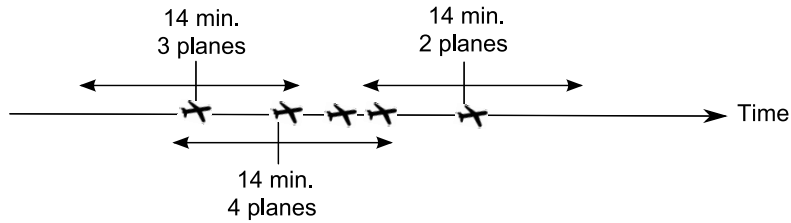}}
\caption{Comparison of time-based methodology and aircraft based
methodology} \label{fig:methodologies}
\end{figure}

\begin{itemize}
\item For each aircraft, the number of planes present in the system
corresponds to the number of airplanes that are, at some point, at
the same time in physically present the TRACON (figure
\ref{fig:aircraftbasedanalysis}).
\item For each aircraft, the flow of entering planes is equal to the number of planes
entering the TRACON during a period of length $T$ centered on the
entrance time of this plane in the TRACON, divided by the period T.
\item For each aircraft, the flow of landing planes is equal to the number of planes
landing at SFO airport during a period of length $T$ centered on the
landing time of this plane at SFO airport, divided by the period T.
\item For each aircraft, the number of rerouted planes that is equal to the number of
planes present in the system that have a \emph{rerouted} trajectory.
\item For each aircraft, the number of runway used corresponds to an
average on a period of length $T$ centered on the plane landing. If
one runway is used more than 75\% of the time during this period, we
will say than only one runway is in use.
\end{itemize}

\subsubsection{Aircraft based analysis of landing and
rerouting}\label{subsubsection:aircraftbasedanalysis} 
\begin{figure}[ht]
\begin{center}
\includegraphics[width = 0.5\textwidth]{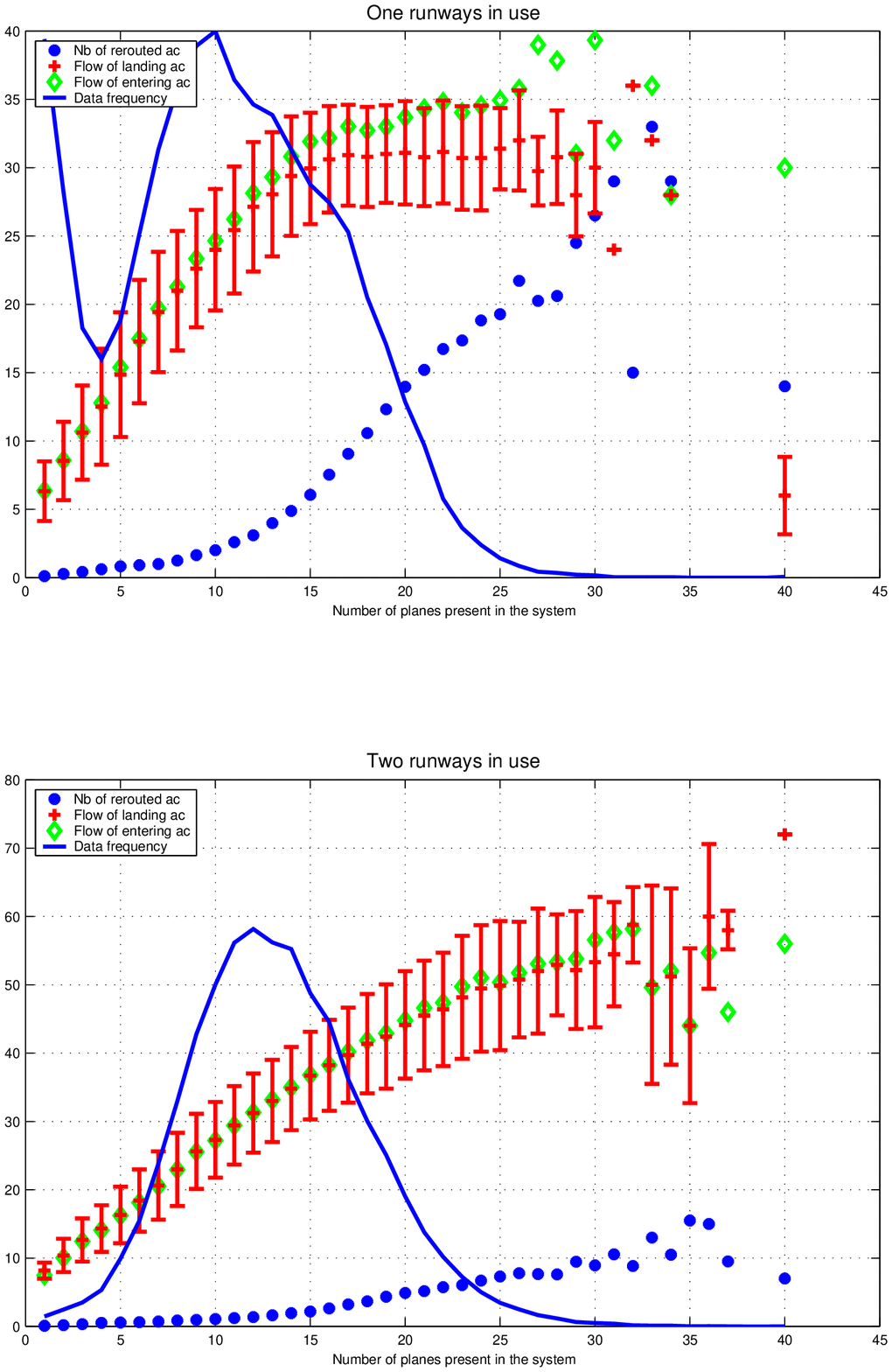}
\caption{Aircraft based analysis of landing and rerouting}
\label{fig:planeBasedAnalysis}
\end{center}
\end{figure}

This section
presents the results of the aircraft based methodology. The figure
\ref{fig:planeBasedAnalysis} plots the average flow of entering
aircrafts in the TRACON (green diamonds), the average flow of
landing aircrafts (red crosses) and its standard deviation (vertical
red lines), and the average number of rerouted aircrafts (blue
stars), as functions of the number of planes in the TRACON and of
the number of runway in use. The definitions of those term are given
in section \ref{sec:aircraft methodology def}. The blue line
represents the data frequency (i.e the number of time this
configuration happened). The period chosen is 15 minutes, to match
with the average time in the system which is 14 minutes 30 seconds.

\begin{description}
\item[One runway in use:] Up to 13 planes present in the TRACON, the flow
of entering and landing aircrafts are very close and proportional to
the number of planes in the system. The average flow of entering
aircrafts is slightly bigger than the average flow of landing
aircrafts. Over 14 planes present in the TRACON, the flow of landing
aircrafts remains at 31 planes an hour whatever the number of
entering aircrafts and the number of planes in the TRACON. The flow
of entering planes goes up to 35 planes per hour. Up to 12 planes
present in the TRACON, the number of rerouted planes increases
slowly. Over 13 planes present in the TRACON, this number increases
quickly and reaches 20 rerouted planes for 25 planes present in the
TRACON. Over 25 planes in the system, the number of rerouted planes
stabilizes around 15. The cases over 25 planes present in the TRACON
are isolated cases (happened once or twice in 9 month). The standard
deviation of the number of landing planes starts at 2.2 aircrafts
and is almost all the time around 4.5 aircrafts.

\item[Two runways in use:] The flow of entering aircrafts and the
flow of landing aircrafts are almost the same up to 29 planes in the
system. The number of landing aircrafts increases almost
proportionally with the number of planes in the TRACON.  Over this
number of planes in the TRACON, the data frequency is very low
(isolated cases) so the results are not very relevant. The number of
rerouted planes increases slowly and reaches at the maximum 10 for
30 planes in the system.

\end{description}

\subsection{Airport characteristics} From the analysis of figures
(\ref{fig:dailyAnalysis}) and (\ref{fig:planeBasedAnalysis}) we can
show characteristics of San Francisco airport. First of all, the
analysis gives a runway capacity a little above 30 planes per hour.
This can be explained by the fact that the second runway can have
been used a few time during the period, increasing the number of
landing on one runway and by the fact that the capacity of 30
aircrafts is a recommandation. To reach the maximum capacity of
landings in one runway configuration, 15 planes in the system are
necessary. When this limit is reached, the number of rerouted planes
increases very fast. The runway occupation has to be optimized and
hence, planes have to be rerouted to arrive at the exact time they
have been assigned by the controllers. When the number of planes
entering the system is high, most of them have to be rerouted.

In the two runway configuration, whatever the number of planes in
the system, the maximum capacity is hardly reached. Hence, the
number of rerouted planes is not high. This configuration can handle
a pick of arrivals or a large number of aircrafts in the TRACON
without having to reroute all the planes.

\subsection{TRACON characteristics}
From the analysis of figures (\ref{fig:dailyAnalysis}) and
(\ref{fig:planeBasedAnalysis}), it is clear that over 15 aircrafts
in the system, the maximum landing capacity is reached. Letting more
planes entering the TRACON leads to reroute them. This increases the
TRACON controller workload, increase the density of aircrafts and
the airspace complexity \cite{keumjinLee} and decreases the degrees
of freedom for rerouting. An certain amount of planes is required to
reach the maximum capacity of landings, but over it, it decreases
the safety and does not improve performances.

\subsection{Pros and cons of both methodologies}
\subsubsection{Time based methodology}

\begin{description}
\item[Pros] This methodology is easy to understand and to interpret.
It is a usual way to analyze data. As shown in figure
\ref{fig:dailyAnalysis} the evolution of the number of rerouted
planes in time is very clear.\newline

\item[Cons] The time period is given and the start of the first
period is arbitrary chosen: the first minute of the first day of
records. As it is arbitrary, some limit phenomenon can appear and
change in a slight way the values calculated. Figure
\ref{fig:methodologies} shows the impact of this arbitrary chosen
time period.
\end{description}

\subsubsection{Aircraft based methodology}
\begin{description}
\item[Pros] this methodology gets rid of the problem of arbitrary chosen
time slot. The planes in the system correspond to the flying planes
that interact which each other: all the planes flying simultaneously
in the TRACON. This has a real meaning when talking about density or
complexity. This method grants a very easy comparison of the results
with a discrete time model. The analysis is exactly the same for all
the aircrafts.

\item[Cons] For the number of planes in the system, we consider the
planes flying in the TRACON. This means we assume that the planes
already landed do not have any more incidence on the planes flying
and neither the planes outside the TRACON.
\end{description}

\section{Model Calibration and Validation}
A model of the TRACON has been developed. One of the aim of
developing such a model is to see if it is possible to reproduce
with a simple model the phenomenon we observed on the real system.
This model presents similarities with the push-back and departure
process model in \cite{pujet}.

\subsection{Description of the model} This model is a discrete time
input-output model. The model has been developed and used with
Matlab.
\begin{description}
\item[System:] The modeled system is the TRACON with a 2 parallels runways
airport. In a nominal configuration (i.e. the 2 runway are open and
simultaneous landing are possible), the maximum landing capacity is
60 aircrafts per hour, 30 for each runway.
\item[Inputs:] The input to the system is the sequence of entering
aircrafts in the TRACON.
\item[Outputs:] The output to the system is the sequence of landing
aircrafts on the two runways.
\item[Time:] The time is divided in 30 seconds slots.
\item[Travel time:] When a plane enters the system, a
\emph{nominal} travel time is randomly drawn with a probability law
describing the configuration of the TRACON (i.e the different
possible routes leading aircrafts from the entrance to the runway
threshold). This nominal time corresponds to the time the plane
would spend in the TRACON if the runway is available without having
to wait.
\item[Runway opening:] To degrade the system, it is possible to
close one or both runways. If a runway is closed none aircraft can
land on it.
\item[Runway availability:] Once a plane lands, the runway is
unavailable for 2 minutes, which is 4 time slot.
\item[Runway attribution:] When a plane enters the system and is
attributed a nominal travel time, it requests the runway for the 2
minutes period corresponding to the end of its travel time. If the
runway is available, it can land and the runway becomes unavailable
for this period. If none runway is available for this period, the
plane has to queue.
\item[Queuing:] If the runway is not available, the plane will wait
until one runway is available. As soon as one runway is available,
the plane lands.
\end{description}

\subsection{Model calibration} 
\begin{figure}[htbp]
\begin{center}
\includegraphics[width = 0.3\textwidth]{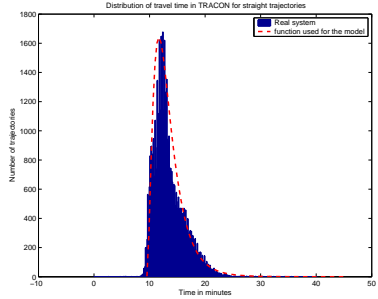}
\caption{Distribution of travel time in the TRACON}
\label{fig:timeInTracon}
\end{center}
\end{figure}

The probability law used to draw
nominal travel times is presented in figure
(\ref{fig:timeInTracon}). In blue, the travel time for straight
trajectories distribution extracted from the data and in red, the
probability law used for the model.

\subsection{Model validation} To validate the results of this model,
the analysis of the data presented in the previous chapter have been
used. From it, the entering sequence of aircrafts has been extracted
and used as input for the model. The number of open runways has been
extrapolated from the landing sequences in the data.

\begin{figure}[htbp]
\begin{center}
\includegraphics[width = 0.7\textwidth]{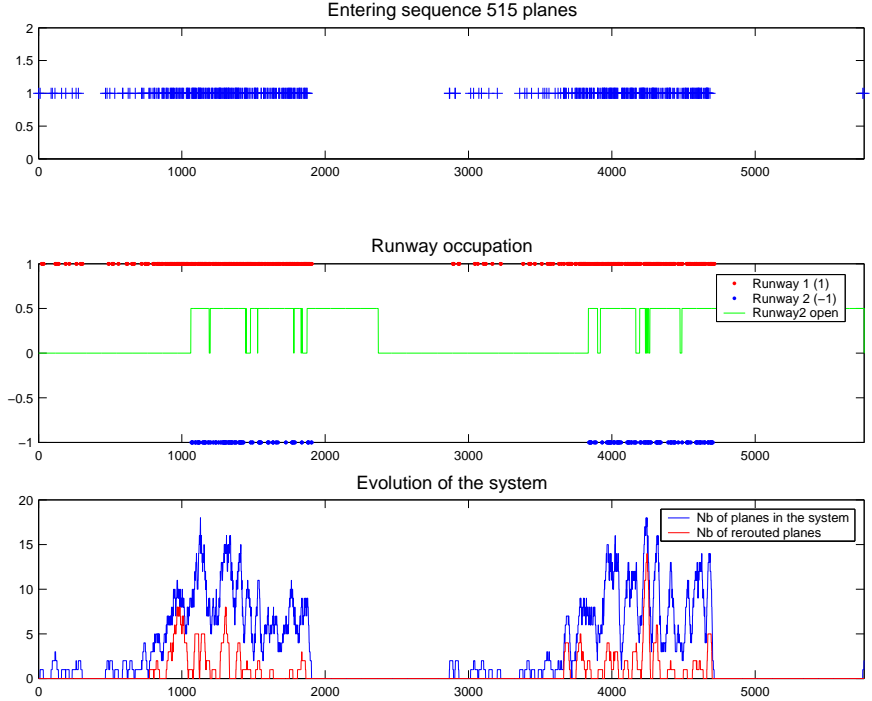}
\caption{Evolution of the modeled system for a two days simulation.}
\label{fig:modelRealisation}
\end{center}
\end{figure}

Figure (\ref{fig:modelRealisation}) presents the evolution of the
modeled system for a two days simulation. The top graph presents the
entering sequence, the second graph presents the runway occupation:
a dot for each time slot when the runway is occupied. The first
runway which is always open is in red with value $1$ and the second
runway has the value $-1$ and is in blue. The green line presents
the period where the second runway is open. The third graph presents
the evolution of the system. Blue line represents the number of
aircrafts in the system and red line represents the number of
rerouted planes. The horizontal axis is the time, in 30 seconds time
slots. One day = 2880 slots.

\begin{figure}[ht]
\begin{center}
\includegraphics[width = 0.5\textwidth]{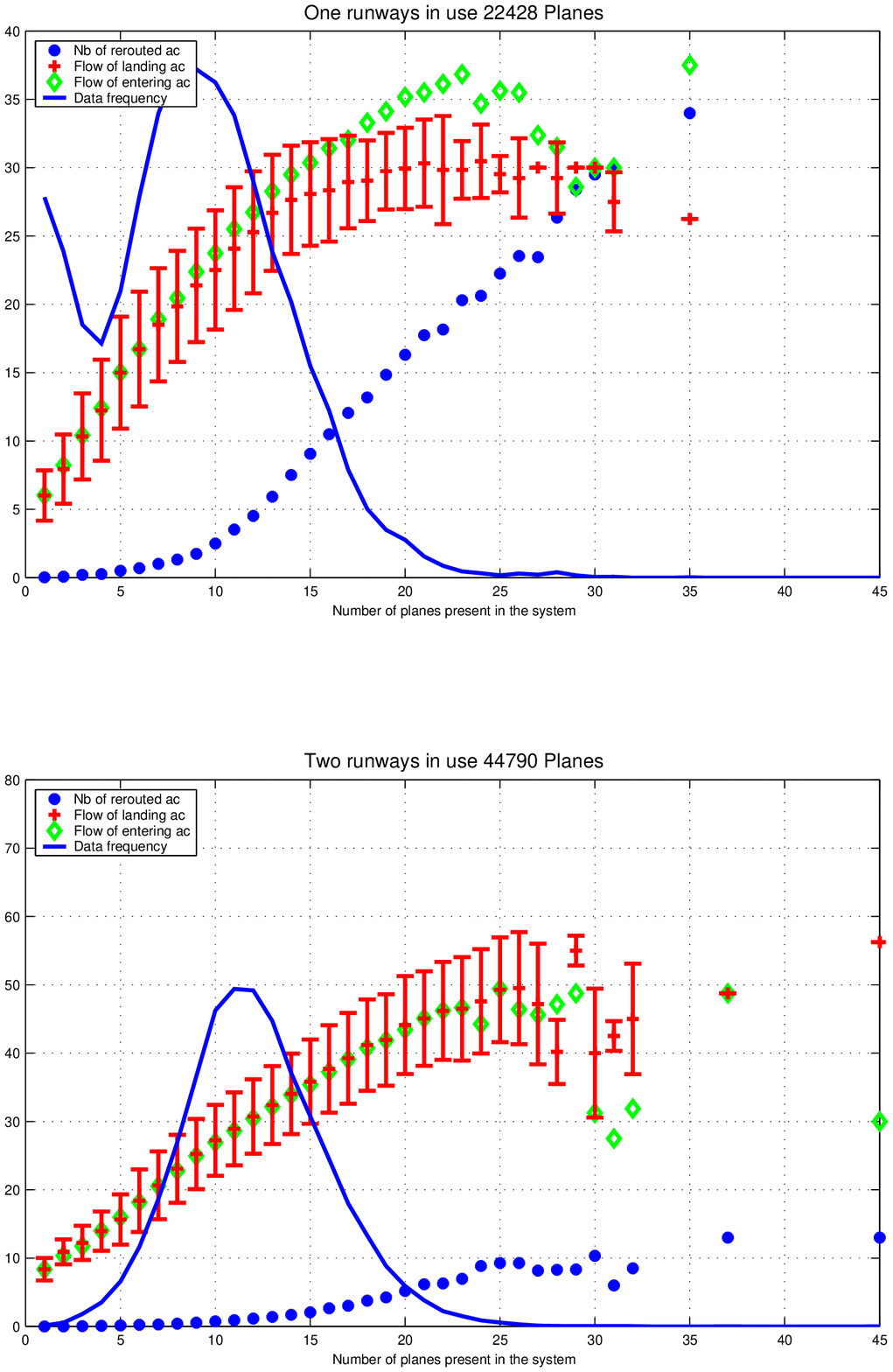}
\caption{Aircraft based analysis of landing and rerouting on the
model} \label{fig:planeBasedAnalysisModel}
\end{center}
\end{figure}

\begin{figure}[ht]
\centering \subfigure[Flow of landing aircrafts, 1
runway]{\label{fig:comparisonLanding1}
    \includegraphics[height = 0.4\textwidth, width=0.4\textwidth]{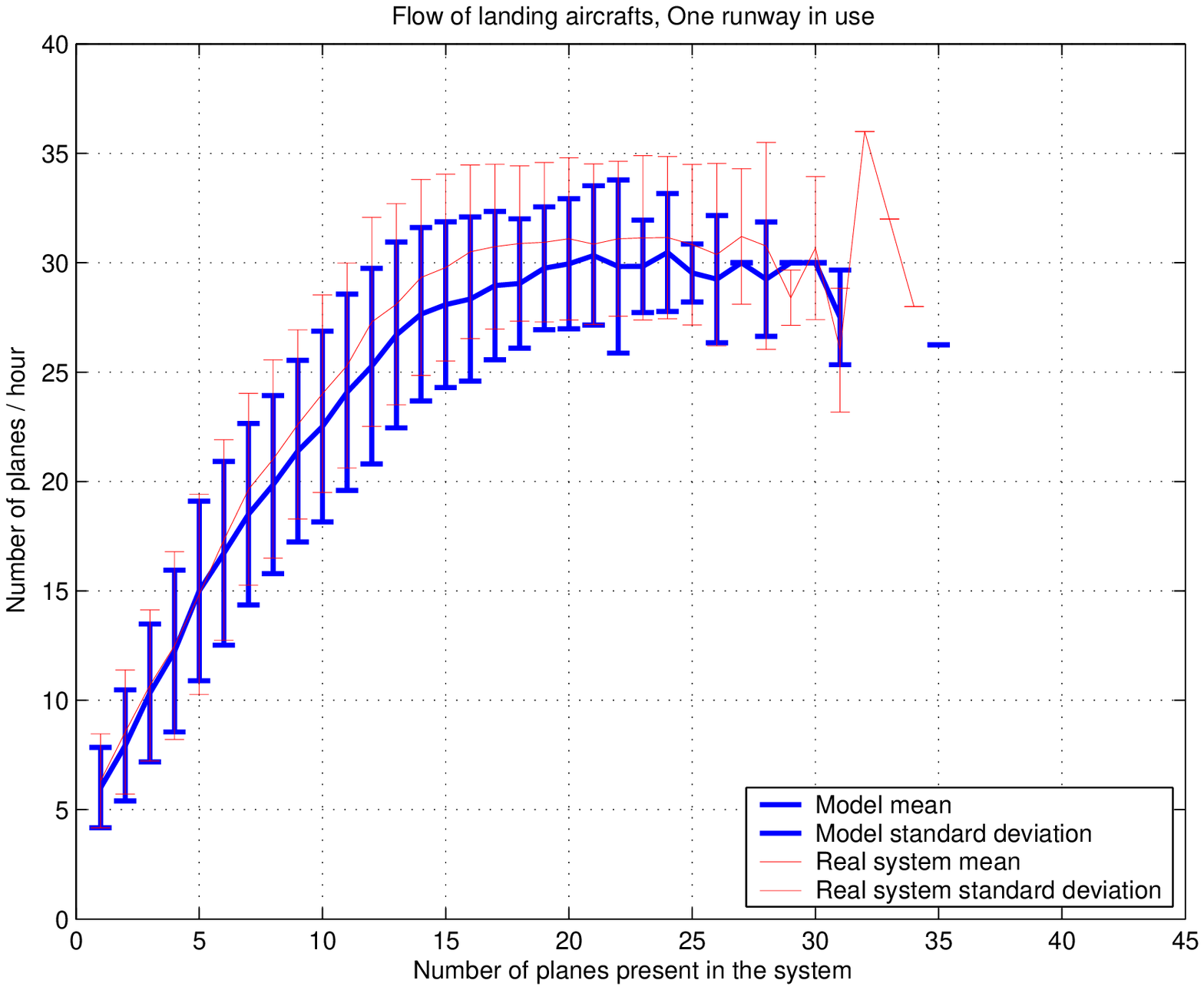}}
\subfigure[Flow of landing aircrafts, 2
runways]{\label{fig:comparisonLanding2}
    \includegraphics[height = 0.4\textwidth, width=0.4\textwidth]{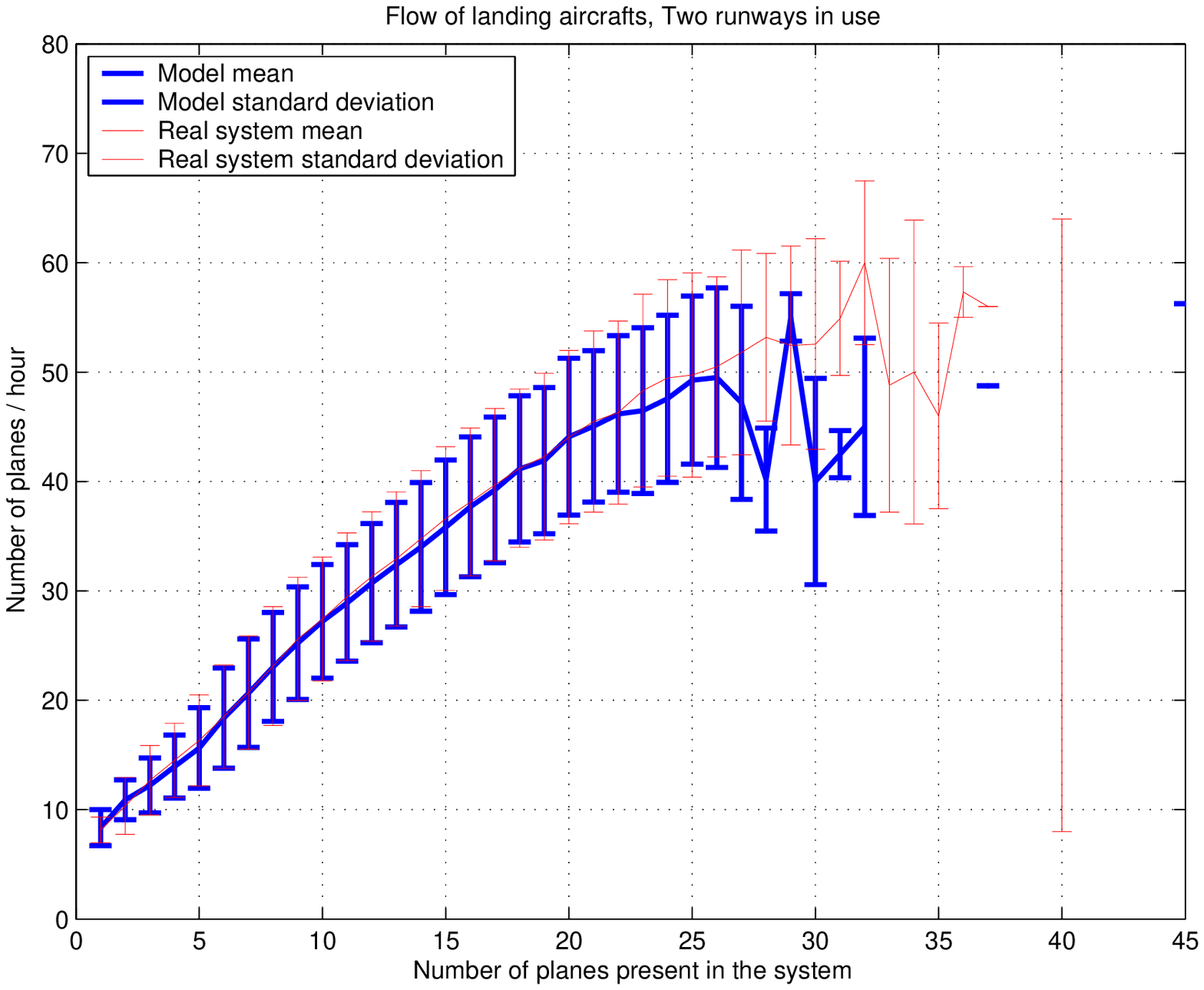}}
    \subfigure[Number of rerouted aircrafts, 1 runway ]{\label{fig:comparisonReroute1}
    \includegraphics[height = 0.4\textwidth, width=0.4\textwidth]{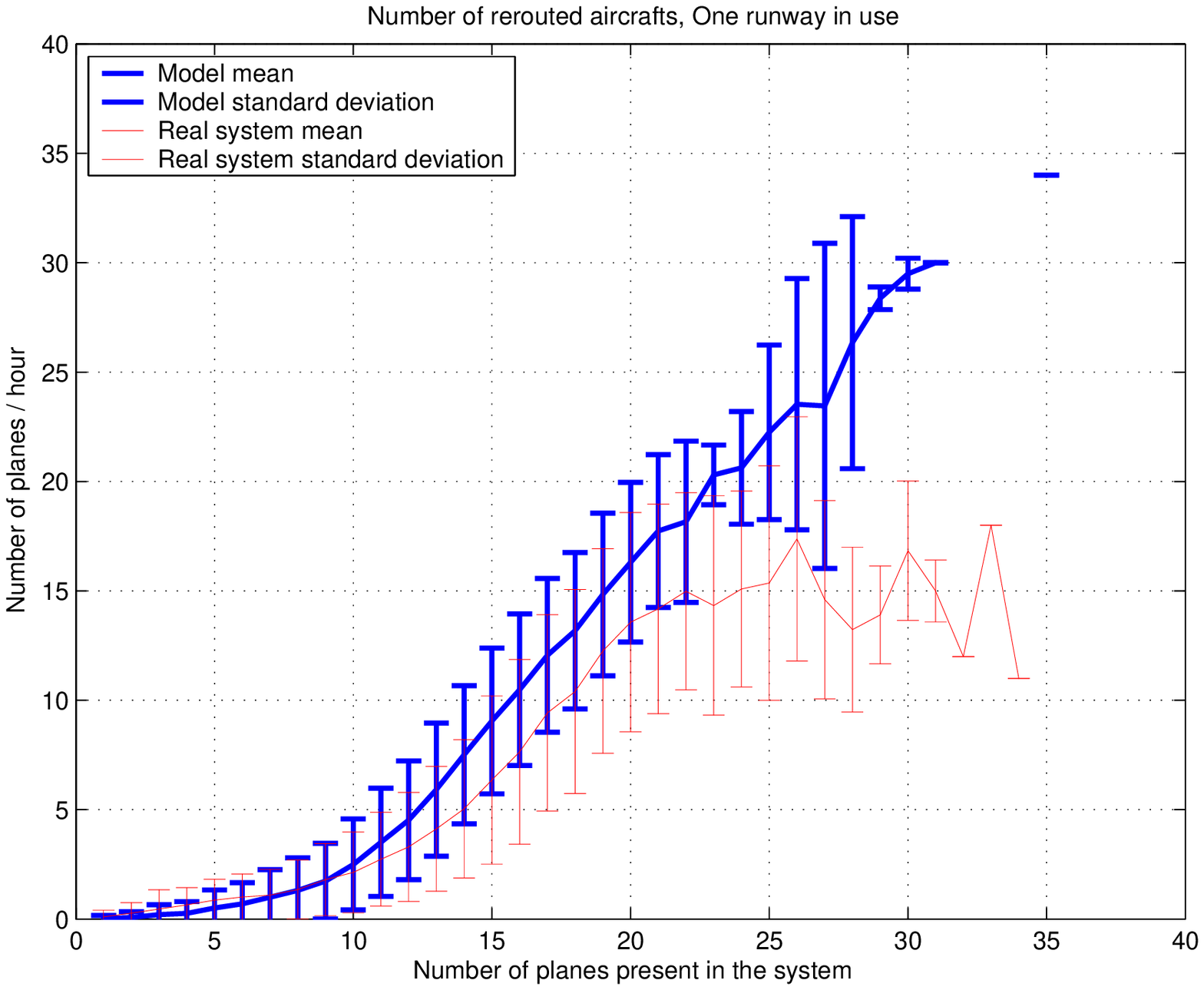}}
\subfigure[Number of rerouted aircrafts, 2 runways
]{\label{fig:comparisonReroute2}
    \includegraphics[height = 0.4\textwidth, width=0.4\textwidth]{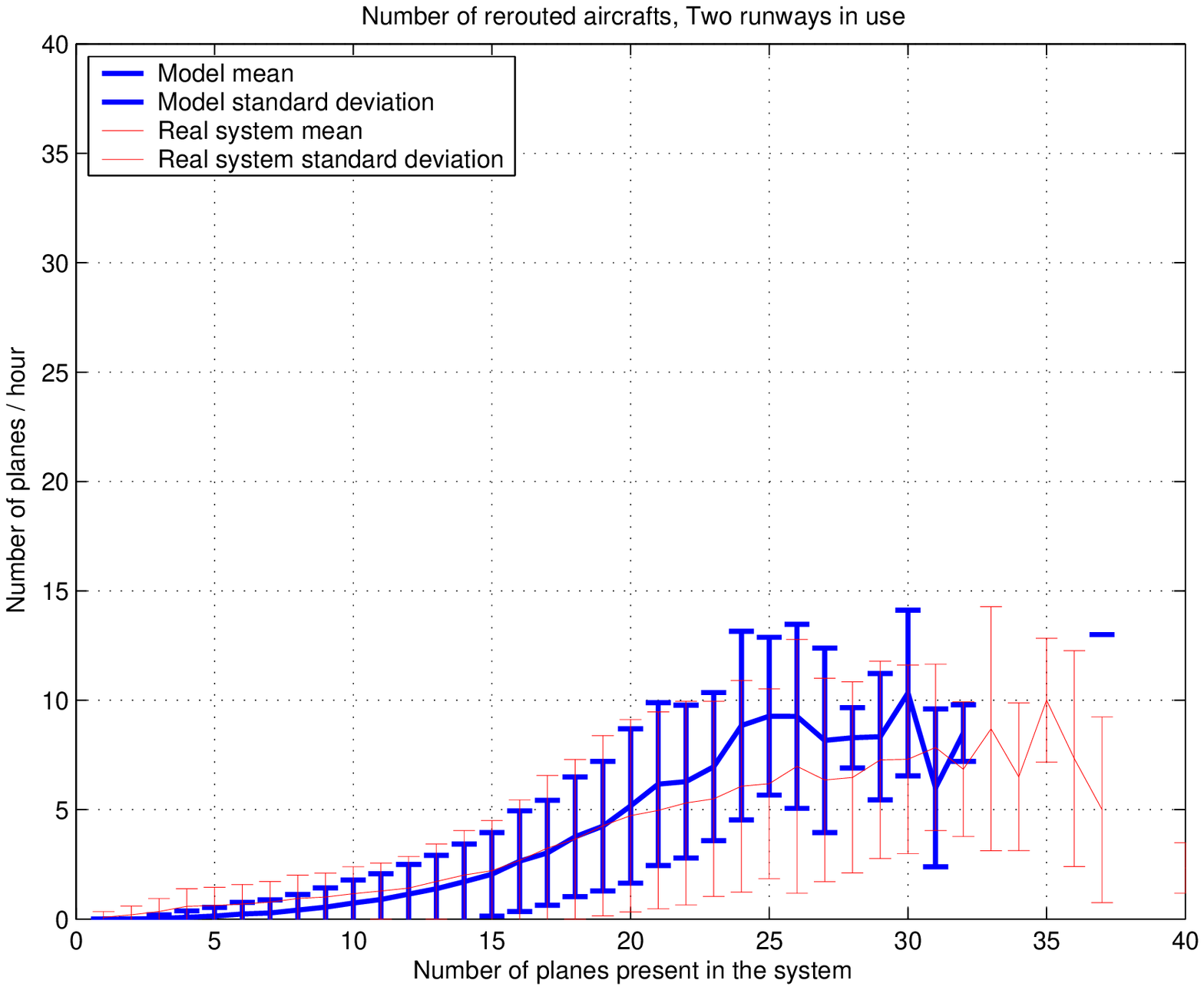}}
\caption{Comparison of landing flow and rerouting between the model
and the real system} \label{fig:comparisonModelReal}
\end{figure}

For figures (\ref{fig:planeBasedAnalysisModel}) and
(\ref{fig:comparisonModelReal}), the model has run on 250 days,
corresponding to 67218 planes, 22428 landing in one runway
configuration and 44790 landing on two runways configuration.

Figure (\ref{fig:planeBasedAnalysisModel}) presents for the model
exactly the same analysis than section
\ref{section:dataPresentation}.\ref{subsection:dataanalysis}.\ref{subsubsection:aircraftbasedanalysis}
. The figure plots the average flow of entering aircrafts in the
TRACON (green diamonds), the average flow of landing aircrafts (red
crosses) and its standard deviation (vertical red lines), and the
average number of rerouted aircrafts (blue stars), as functions of
the number of planes in the TRACON and of the number of runway in
use. The definitions of those term are given in section
\ref{sec:aircraft methodology def}. The blue line represents the
data frequency. For the model, the period chosen is 16 minutes to
catch the real flow of this discrete model.

To compare the results, figure (\ref{fig:comparisonModelReal})
presents the comparison between the model and the real system.
Figures (\ref{fig:comparisonLanding1}) and
(\ref{fig:comparisonLanding2}) present the comparison of the landing
flow between the model and the real system. The blue line represents
the average flow of landing aircraft in function of the number of
planes in the system for the model and the red line is for the real
system. The vertical lines represent the standard deviation. Figures
(\ref{fig:comparisonReroute1}) and (\ref{fig:comparisonReroute2})
present the comparison of the landing flow between the model and the
real system. The blue line represents the average flow of landing
aircraft in function of the number of planes in the system for the
model and the red line is for the real system. The vertical lines
represent the standard deviation.

As shown in figures (\ref{fig:comparisonLanding1}) and
(\ref{fig:comparisonLanding2}), the behavior of the model concerning
the flow of landing aircrafts given by the model is very close to
the real system. When there is one runway in use, the model predicts
a flow a little lower than in the real system. For high numbers of
planes in the system, the average flow is 30 landings/hour which
corresponds to the specification. Notice that the standard deviation
of the model and the real system are always very close. When there
are two runways in use, up to 25 planes in the system, the model
predicts exactly the same behavior than the real system. Over 25,
there are some discrepancies but the number of realization is very
small. \newline

On one runway configuration, figure (\ref{fig:comparisonReroute1})
shows that the predicted number of rerouted plane is pretty accurate
up to 10 planes in the system. Between 10 and 22 planes in the
system, prediction are slightly higher than the real system. Over 22
planes in the system, the prediction keeps increasing at the same
rate while in the real system, it stabilizes. On the two runways
configuration, figure (\ref{fig:comparisonReroute2}) shows that the
prediction are accurate up to 20 planes in the system and over 20,
slightly higher.

\subsection{Analysis of the model results} The maximum number of
planes present in the system is greater in the real system than in
the model. An explanation could be that there is always one runway
open in the model contrary to the real system where sometimes, both
runway can be closed. On one runway configuration, the model
predicts an always increasing number of rerouted planes with the
number of planes in the system. This seems logical, while in the
real system, it stops increasing.

\subsection{Perspective of evolution}
\begin{itemize}
\item Once a plane lands, the unavailability time of the runway is
always exactly two minutes. This does not take in account the size
of the aircraft. In a future model, the unavailability time could be
dependant of the size of the aircraft. The size of the aircraft
would be randomly drawn with a probability law based on the real
distribution of aircrafts. A distribution in three kind of aircrafts
is used in \cite{pujet}: medium jets, large jets and heavy jets.
\item The queuing is very deterministic. When a plane is queuing, as soon as the runway is
available, the plane will land. It does not take in account the fact
that planes have to keep moving and hence, if it is crowded, it can
be difficult to arrive with such a accuracy. An small uncertainty
could be added on the moment a plane queuing lands.
\item The model does not include different routes in the TRACON.
The time spend in the TRACON is randomly drawn with a certain
probability law. This could be more representative if the entering
sequence would include several entry points corresponding to
different routes with different travel time probability law
different for each route. Some route could be longer or more
frequent.
\end{itemize}

\section{Applications}
\subsection{Graceful degradation}
An application for the model presented above is graceful degradation
for air traffic control systems. To define \emph{graceful
degradation}, we introduce two systems (1) a nominal system (2) a
degraded system. \emph{Graceful degradation} happens when the
transition from the nominal system to the degraded system is always
smooth and with no event.

Degradation can occur for many reasons such as bad weather,
breakdown of control systems (e.g radar) or communication systems...
In our case, the nominal system is given by the TRACON, San
Francisco airport with the two runways in use and the air traffic
controllers managing the plane arrivals. The degraded system is
still the TRACON but with only one or zero runway in use, and/or the
controllers not able to manage the system for some reason. This
degraded system must be sustainable. The current way to get a
\emph{graceful degradation} of this system is to reroute planes (i.e
to modify their trajectories so that they can safely wait before
landing) or even some ground slops for short haul traffic (e.g
United shuttle from LAX to SFO). The interesting fact about San
Francisco airport is that this phenomenon is tightly related to the
weather. As the two runways are very close, as soon as the
visibility is bad, which occurs very often due to the fog in San
Francisco, it is impossible to use them at the same time. The system
is hence very often degraded. The complexity \cite{keumjinLee} of
the system must remain under a certain level so that if a
degradation is to happen, the safety is still assured. A way to
reduce complexity is to limit the number of aircrafts present in the
system.

\subsection{Decongestion of the TRACON}
A study about the impact of the landing demand on landing capacity
is done in \cite{McTDMABenefits} for Newark Airport. This study
shows that when demand is higher than capacity of the runway, it
results in a drop of runway capacity. Plausible factors can be the
application of traffic flow restrictions and holding. Other factors
may include a high workload of air traffic controllers and airspace
complexity constraints.
\newline To see the impact of a congested TRACON, arrivals
simulation have been made using San Francisco demand sequence. The
TRACON has been given a limit capacity : no plane can enter it if
this limit is reached. As soon as a plane lands, one plane waiting
out of the TRACON can enter. \emph{Waiting} time refers to the time
the plane were asked to wait outside the TRACON and \emph{rerouting}
time corresponds to the time the planes were asked to wait for the
runway to be available to land. The queue outside the TRACON is a
First-In First-Out queue. The simulations cover 243 days and 66,067
aircrafts.

\begin{figure}[ht]
\centering
       \subfigure[Mean rerouting and waiting time]
            {\label{fig:delayComparisonMean}
            \includegraphics[height = 0.4\textwidth, width=0.4\textwidth]{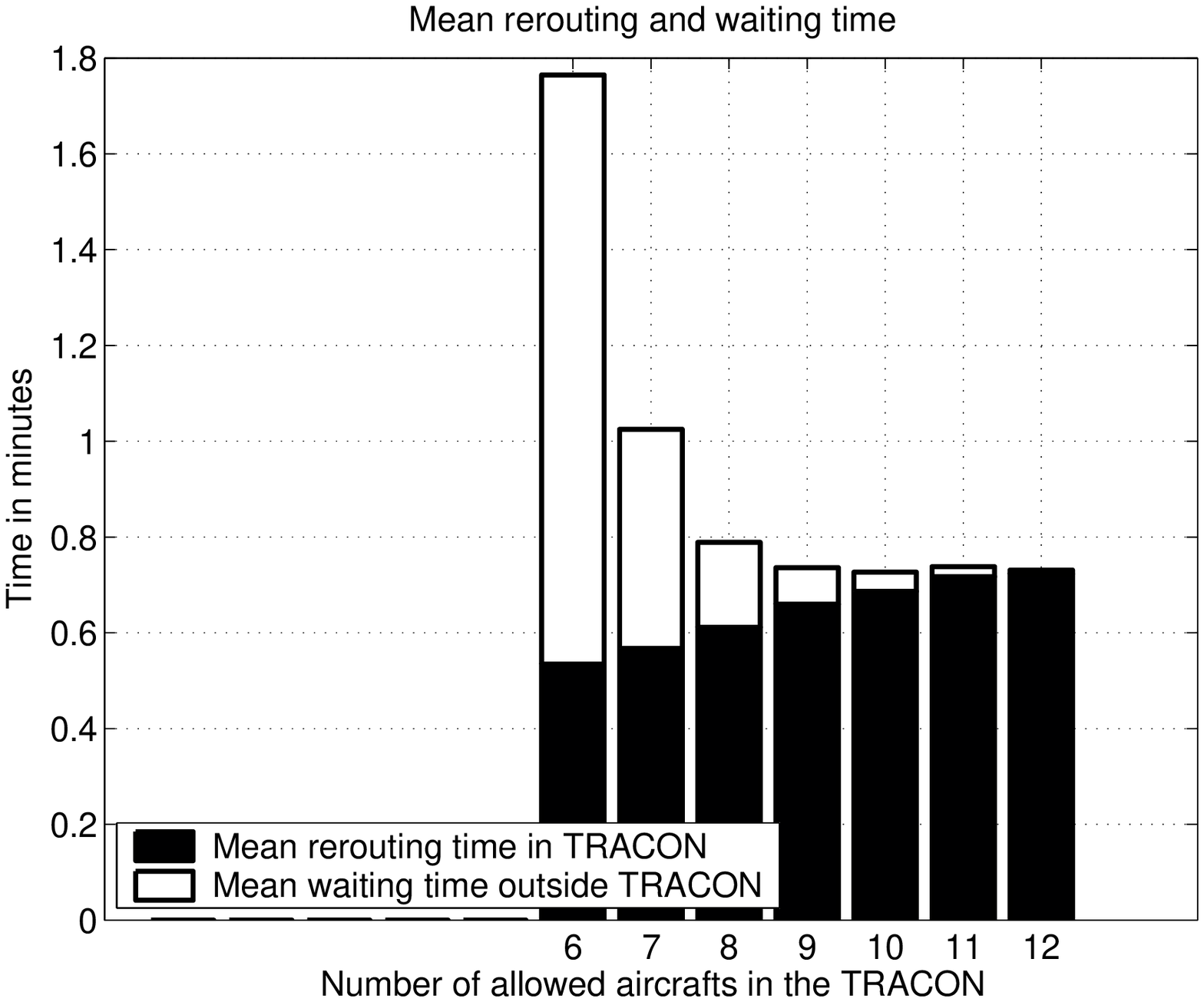}}
    \subfigure[Percentage of delayed flights]
            {\label{fig:delayomparisonPercent}
            \includegraphics[height = 0.4\textwidth, width=0.4\textwidth]{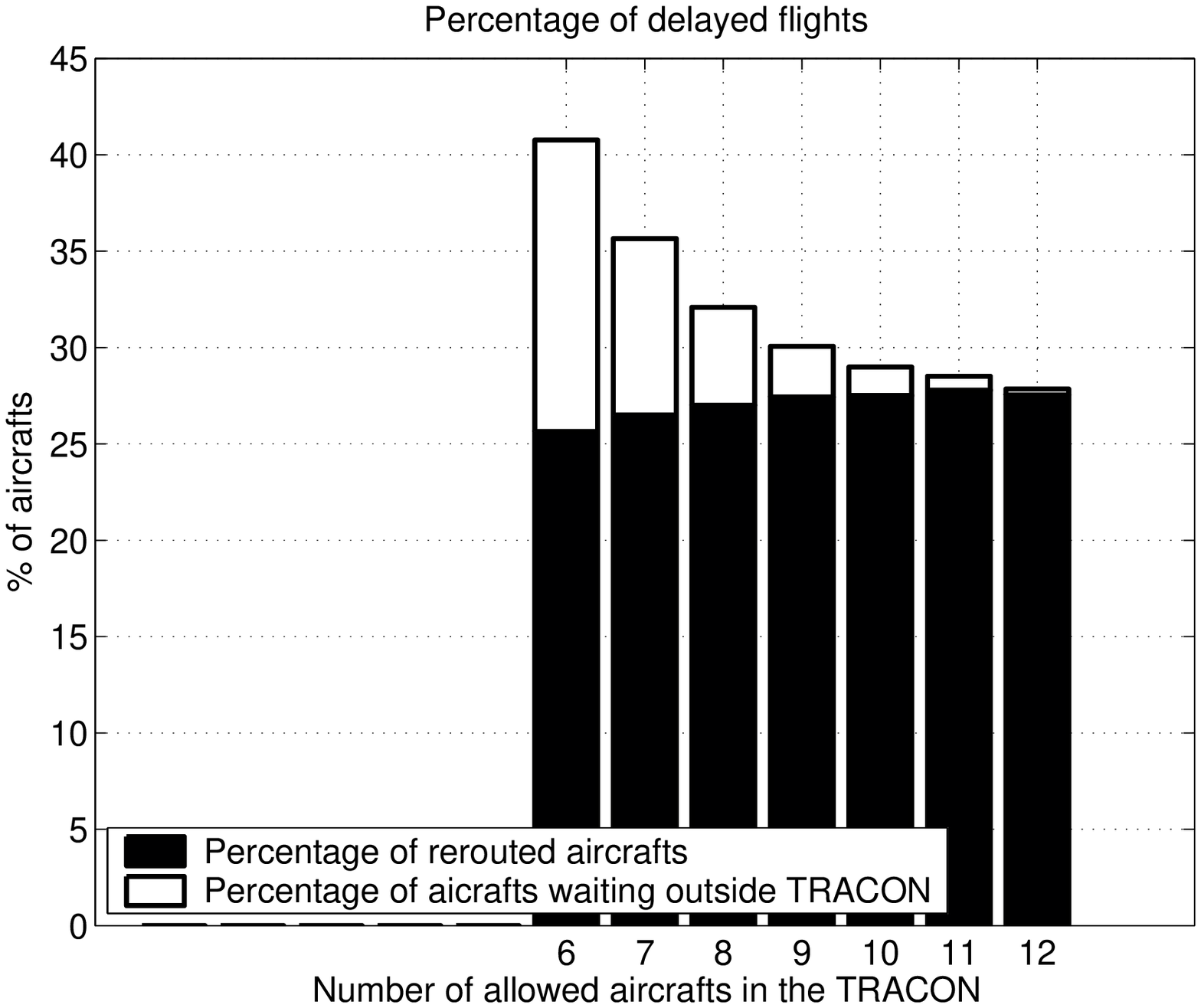}}
    \caption{Simulated delays function of the number of allowed
    planes in the TRACON }\label{fig:delayComparison}
\end{figure}

Figure \ref{fig:delayComparison} presents the results of simulation
for TRACON limit capacity varying from 6 to 12 aircrafts. Black bars
stand for rerouting values and white bars for waiting waiting.
Figure \ref{fig:delayComparisonMean} presents the average delay and
figure presents the  percentage of delayed aircrafts
\ref{fig:delayomparisonPercent}.

A too low TRACON limit capacity has two effects. First, the number
of planes waiting outside the TRACON is very high and second, the
runway capacity will not be reached which results in an inefficient
configuration. Then, when the TRACON capacity increases, the mean
rerouting time decreases and stabilizes while the percentage of
delayed flight slightly decreases. From 9 to 12 planes in the
TRACON, the average rerouting time is almost the same, but, as the
number aircrafts in the TRACON is smaller, the complexity of the
system is smaller. The average rerouting time does not change much
because the runway maximum capacity is reached and planes have to
queue anyway. As shown in the first study, over a certain number of
planes in the TRACON, the number of rerouted aircrafts increases
very quickly. If the number of simultaneous rerouted planes is high,
the controller workload increases. By reducing this workload, the
controller could focus on fewer aircrafts and optimize better the
trajectories which would result in an increase of the runway
capacity. Moreover, the higher the number of present planes in a
given area is, the less is the flexibility for an efficient
rerouting process. TRACON are very congested area and limiting the
number of aircraft present in those areas would increase the safety
without generating extra delays.

\section{Conclusion}
The analysis of San Francisco bay's TRACON records shows the landing
capacity of the airport and the influence of the number of plane
present simultaneously in the TRACON on the rerouting process and
landing capacity. It is not useful to allow a large number of
aircrafts simultaneously in the TRACON to increase performances. A
validated TRACON input-output model is proposed, calibrated and
validated. It is used to show that limiting the number of aircrafts
in the TRACON does not increase the average delay, reduces the  the
controller workload and increases safety.

\section*{Acknowledgement}
\addcontentsline{toc}{chapter}{Acknowledgement}
 This research was supported by THALES. The author wish to thank Bert
Ganoug from the Aircraft Noise Abatement Office of San Francisco
International Airport pleasantly for providing the records of the
TRACON of San Francisco bay.


\begin{thebibliography}{15}

\bibitem{CTAS}
    NASA, \it{Center TRACON Automation System}, http://www.ctas.arc.nasa.gov/project\_description/index.html\#overview.

\bibitem{GDP}
    Michael O. Bally and Guglielmo Lulli, \it{Ground Delay Programs: Optimizing over the Included Flight Set Based on
    Distance}, 2001.

\bibitem{TMA}
    NASA, \it{Center TRACON Automation System, Traffic Management Advisor}, http://www.ctas.arc.nasa.gov/project\_des- cription/tma.html.


\bibitem{McTMA}
    NASA, \it{Center TRACON Automation System,  Multi-Center Traffic Management Advisor},
    http://www.ctas.arc.nasa.gov/project\_description/mctma.html.



\bibitem{McTMAresults}
    Todd C. Farley, Steven J. Landry, Ty Hoang, Monicarol Nickelson, Kerry M. Levin, Dr. Dennis Rowe and Dr. Jerry D. Welch,
    \it{Multi-Center Traffic Management Advisor: Operational Test Results},
    AIAA 5th Aviation, Technology, Integration, and Operations Conference (ATIO), 2005.

\bibitem{FAST}
    NASA, \it{Center TRACON Automation System,  Final Approach Spacing Tool},
    http://http://www.ctas.arc.nasa.gov/project\_description/fast.html.

\bibitem{Ioannis}
    Ioannis Anagnostakis, \it{A Multi-Objective, Decomposition-Based Design
Methodology and its Application to Runway Operations Planning}, PhD
thesis, September 2004

\bibitem{pujet}
    Nicolas Pujet, Bernard Delcaire, Eric Feron, \it{Input-Output Modeling and Control of the Departure Process of Congested Airports},
    1999.



\bibitem{McTDMABenefits}
    Husni Idris and Antony Evans,
    \it{Benefits assessment of multi-center traffic management advisor for Philadelphia and New York}, AIAA Guidance, Navigation, and Control Conference and Exhibit, 2003.

\bibitem{PHLtimebasedmetering}
    Todd Farley, John D. Foster, Ty Hoang and Katharine K. Lee,
    {A time-based approach to metering  arrival traffic to Philadelphia},
    AIAA 2001-5241, 2001.

\bibitem{keumjinLee}
    Keumjin Lee,Eric Feron and Amy Pritchett, \it{Air traffic complexity: An Input-Output approach},
    American Control Conference, New York 2007. To be published, 2007.


\end{thebibliography}
\end{document}